\documentclass[aps,prl,twocolumn,showpacs,superscriptaddress]{revtex4-2}
\usepackage{latexsym}
\usepackage{amssymb}
\usepackage{graphicx}
\usepackage{amsmath}
\usepackage{bm}
\usepackage[colorlinks,
linkcolor=black,
citecolor=black,
urlcolor=blue
]{hyperref}
\usepackage{verbatim}
\usepackage{mathrsfs}
\usepackage{extarrows}
\usepackage{comment}
\usepackage{mathtools,slashed}

\usepackage{cancel}
\usepackage{subfigure}
\usepackage{graphicx}
\usepackage{caption}
\usepackage{ragged2e}
\usepackage{braket}
\usepackage{float}
\usepackage{booktabs}
\usepackage{multirow}

\begin{document}
	
	\title{Twisting-operator approach to identifying gaplessness in SU($N$) fermionic systems}
	
	\author{Hao Cheng}
	\affiliation{School of Physics and Astronomy, Shanghai Jiao Tong University, Shanghai 200240, China}
	\author{Hang Su}
	\affiliation{School of Physics and Astronomy, Shanghai Jiao Tong University, Shanghai 200240, China}
	\author{Yuan Yao}
	\email{smartyao@sjtu.edu.cn}
	\affiliation{School of Physics and Astronomy, Shanghai Jiao Tong University, Shanghai 200240, China}
	
	\begin{abstract}
		We propose a general necessary condition for a spinful fermion chain with SU(2) spin-rotation symmetry to be gapped.  
		Specifically, we prove that the expectation value of a properly defined fermionic twisting operator asymptotically approaches unity in any gapped phase with finite ground-state degeneracy, with finite-size corrections bounded by $\mathscr{O}(1/L)$.  
		Consequently, a non-unity value in the thermodynamic limit provides a sufficient criterion for identifying gapless fermion chains.  
		We confirm this criterion using the $s$-wave Bardeen-Cooper-Schrieffer (BCS) Hamiltonian and determinant quantum Monte Carlo (DQMC) simulations of interacting Hubbard models. 
		We further extend the twisting-operator approach to SU($N$)-symmetric fermionic systems, where the gapped ground-state sector must be SU($N$)-singlet and $\langle \hat{\mathcal{U}}^M\rangle=1+\mathscr{O}(1/L)$ by a fermionic twisting operator $\hat{\mathcal{U}}$ with a suitably chosen integer $M$.
	\end{abstract}
	
	\date{\today}
	\maketitle
	
	\paragraph{Introduction.---}
	Identifying the phases of different quantum many-body systems has always been an important topic in the field of condensed matter~\cite{Landau:1937aa}.
	The quantum phases can be roughly classified into gapped and gapless phases depending on the low-energy spectra \textit{in the thermodynamic limit}~\cite{Zeng:2015aa,zeng2019quantum}.
	Therefore,
	the recognization of the spectral gap or ground-state degeneracy is difficult due to the complicated interactions and the infinitely large system sizes.
	Nevertheless,
	the symmetry has played an important role in the spectral-gap constraint.
	Lieb-Schultz-Mattis (LSM) theorem and its extension~\cite{Lieb:1961aa,OYA1997,Oshikawa:2000aa,Hastings:2004ab,NachtergaeleSims} state that a SO(3)-symmetric spin chain with half-integer spin cannot have a unique ground state in the presence of lattice translation symmetry with half-integer spin(s) per unit cell. 
	However,
	LSM theorem cannot further distinguish gaplessness and nontrivial gapped ground-state degeneracy and it does not have predictions for integer spin chains.
	
	Recently,
	several gaplessness criteria is proposed by the following twisting operator~\cite{Resta:1998aa,Oshikawa:2000aa,Resta:1999aa,Aligia:1999aa,Lieb:1961aa,Tasaki:2018aa,Su:2024aa,29j7-ltfn} in spin chains,
 	\begin{eqnarray}
 		\hat{U}_\text{spin}= \exp \left( \frac{2 \pi i}{L} \sum_{j = 1}^{L} j \hat{S}^z_j \right),
 	\end{eqnarray}
	where it is shown that $\langle\hat{U}_\text{spin}^2\rangle=1+\mathscr{O}(1/L)\rightarrow1$ for any ground state of a gapped SO(3)-symmetric spin chains.

Many spin models are low-energy effective theory of the underlying electronic systems in the insulating phase with certain filling conditions.
It is a natural question to generalize the gaplessness criteria to the parent fermionic systems which is more fundamental than the SU(2) spin degrees of freedom.
Moreover,
SU($N$) ``spin'' models are also realized in ultracold atoms on optical lattices~\cite{Wu:2003aa,Honerkamp:2004aa,Cazalilla:2009aa,Gorshkov:2010aa,Taie:2012aa,Pagano:2014aa,Scazza:2014aa,Zhang:2014aa},
so the extension to $N$-``flavor'' fermionic systems is also of practical interest for $N>2$. 

In this \textit{Letter}, we extend those twisting-operator criteria in two directions.  
	First, we generalize it to spinful fermion chains by defining a fermionic twisting operator $\hat{U}$~\cite{PhysRevResearch.2.023277,PhysRevB.107.075153} with the local fermion spin operator $\hat{S}^z_{f,j}$ in Eq.~\eqref{ferm_spin}.  
	We show that, for any gapped fermionic Hamiltonian with finite ground-state degeneracy, $\langle \hat U^2\rangle=1+\mathscr{O}(1/L)\rightarrow1$, and confirm this result using the BCS Hamiltonian~\cite{PhysRev.106.162,Altland_Simons_2010} and DQMC simulations~\cite{blankenbecler1981monte,scalapino1981monte,assaad2008world,SmoQyDQMC.jl} of interacting Hubbard models. 
	Second, we extend the argument to SU($N$)-symmetric fermionic chains~\cite{PhysRevB.86.235142,PhysRevB.75.184441,andp.20085201204,PhysRevB.91.174427,tanimoto2015,PhysRevB.93.155134,PhysRevB.98.085104,PhysRevB.101.195121,Affleck1986} and prove that the gapped ground-state sector must be SU($N$)-trivial, with $\langle \hat{\mathcal{U}}^M\rangle=1+\mathscr{O}(1/L)\rightarrow1$ for a series of $\hat{\mathcal{U}}$ with its corresponding $M$.

	\paragraph{Fermionic local Hilbert space.---}
	We consider a spin-1/2 fermion chain of length $L$. The local Hilbert space on each site $j$ is four dimensional,
	$\mathscr{H}_j=\mathrm{span}\{|0\rangle_j,|\uparrow\rangle_j,|\downarrow\rangle_j, |\uparrow\downarrow\rangle_j\}$, 
	where the empty state $|0\rangle_j$, and the doubly occupied state $|\uparrow\downarrow\rangle_j=c^\dagger_{j\uparrow}c^\dagger_{j\downarrow}|0\rangle_j$ are spin singlets, 
	while $|\uparrow\rangle_j$ and $|\downarrow\rangle_j$ form the fundamental spin-1/2 representation of SU(2). 
	Therefore the local spin operator may be written as
	\begin{eqnarray}\label{ferm_spin}
		\hat{S}^z_{f,j} = \frac{1}{2}(c_{j,\uparrow}^{\dagger}c_{j,\uparrow }-c_{j,\downarrow}^{\dagger}c_{j,\downarrow }),
	\end{eqnarray} 
	and the fermion operators satisfy $\{c_{i,\alpha},c_{j,\beta}^{\dagger}\} = \delta_{ij}\delta_{\alpha\beta}$. 
	Equivalently, in the basis $\{|0\rangle,|\uparrow\rangle,|\downarrow\rangle,|\uparrow\downarrow\rangle\}$, 
	one has $\hat{S}^z_f = \frac{1}{2} \mathrm{diag} (0, 1, -1, 0)$.
	Thus the fermionic Hilbert space decomposes into a spin-1/2 doublet and two SU(2)-singlet sectors.
	
	\paragraph{Symmetry and locality.---}
	To set the stage, 
	we introduce various notations and concepts that are essential in our discussion of spin chains.
	A Fermion-chain Hamiltonian $\mathcal{H}$ is defined on a one-dimensional lattice of length $L$,
	where $\hat{S}^z_{f,j}$ acts on the Hilbert space on site $j$, under periodic boundary conditions (PBCs), $\hat{S}^z_{f,j+L}=\hat{S}^z_{f,j}$. The Hamiltonian possesses SU(2) spin-rotation symmetry;
	
	The Hamiltonian is assumed to be \textit{local}, 
	i.e., it admits a decomposition  
	$\mathcal{H}=\sum_{j=1}^{L}h_j$,
	where each $h_j$ acts as the identity operator on the spins at a distance larger than $d_L$ from site $j$,
	with $d_L$ being upper bounded by some model-dependent $d$ for $L\gg1$.  
	Moreover, the local decomposition can always be chosen to respect the on-site symmetry~\cite{Su:2024aa},
	\begin{eqnarray}\label{local_sym}
		[\hat{S}^z_f,h_j]=[R,h_j]=0,
	\end{eqnarray}
	where $R$ is any group element of SU(2).
	
	Furthermore, the Hamiltonian is assumed to be invariant under a translation 
	$\hat{T}^{-1}_r \hat{S}^z_{f,j} \hat{T}_r = \hat{S}^z_{f,j+r}$,
	where an integer $r$ is a model-dependent minimum period,
	so that we can increase the system size $L$ as $L=Nr$ with $N\rightarrow\infty$ to define the thermodynamic limit.


	\paragraph{Twisting operator for Fermion.---}
	Since $S^z$ and $S^z_f$ follow the same commutation relation, we can directly instead $S^z$ by $S^z_f$ to obtain the corresponding twisting operator
	\begin{eqnarray}
		\hat{U}= \exp \left( \frac{4\pi i}{L} \sum_{j = 1}^{L} j \hat{S}^z_{f,j} \right).
	\end{eqnarray}
	There is a useful relationship between $\hat{U}$ and $\hat{T}_r$: 
	\begin{eqnarray}\label{TUT}
		\hat{T}^{-1}_r \hat{U} \hat{T}_r=\hat{U} 
		\underbrace{\exp\!\left(4\pi i \sum_{m = 1}^{r} \hat{S}_{f,m} \right)}_{\text{phase}} 
		\underbrace{\exp\!\left(- \frac{4\pi i}{L}  r\hat{S}^z_{f,\mathrm{tot}} \right)}_{\text{global rotation}}, 
	\end{eqnarray}
	where $\hat{S}^z_{f,\mathrm{tot}} = \sum_{m=1}^{L}\hat{S}^z_{f,m}$ and $4\pi$ (rather than the conventional $2\pi$) is chosen to ensure $\exp\!\left(4\pi i \sum_{m = 1}^{r} \hat{S}_{f,m} \right)$ is proportional to identity and simply a phase.
	By using this twisting operator, the conclusions and proofs we obtain in the spin system can be directly extended to the fermion system.
The phase and the global rotation property in Eq.~\eqref{TUT} sufficiently give the following theorem~\cite{Su:2024aa}: 
	

	\textit{Theorem~1: Singlet criterion.---}
	Given a SU(2)-symmetric gapped fermionic chain Hamiltonian $\mathcal{H}$ with a finite ground-state degeneracy,
	any of its ground state(s) form a trivial representation(s) of SU($2$).
	
	Additionally, we can define $\hat{S}^x_{f,j}\equiv \frac{1}{2}\left(c^{\dagger}_{j,\downarrow}c_{j,\uparrow}+c^{\dagger}_{j,\uparrow}c_{j,\downarrow }\right)$
which satisfies that 
\begin{eqnarray}
R^{\dagger}\hat{S}^z_{f,i}R + \hat{S}^z_{f,i} = 0,
\end{eqnarray}
	where $R=\exp(i\pi \sum_{j}\hat{S}^x_{f,j})$.
	It can be shown that
	the \textit{existence} of such an $R$ to satisfy the above relation together with \textit{Theorem~1} can produce:
	\paragraph{Theorem~2: $\langle\hat{U}^2\rangle$-criterion.---} 
	Given a SU(2)-symmetric {fermion-chain} Hamiltonian $\mathcal{H}$,
	if any of its ground state(s) $|\text{gs}\rangle$,
	satisfies that
	\begin{eqnarray}
		\lim_{L\rightarrow\infty}\langle\text{gs}|\hat{U}^2|\text{gs}\rangle\neq 1 + \mathscr{O}(1/L),
	\end{eqnarray}
	then the Hamiltonian must either be gapless or have an infinite ground-state degeneracy.
	


	\paragraph{Analytic and numerical illustrations---}
	We now illustrate the fermionic criterion with both an exactly solvable mean-field example and two interacting fermionic models.  
	As a first example, we consider the one-dimensional $s$-wave BCS Hamiltonian~\cite{PhysRev.106.162,Altland_Simons_2010}. 
	Its ground state can be written as
	\begin{eqnarray}
		|\text{G.S.}\rangle = \prod_{p}\left(u_{p}+v_{p}c_{p\uparrow}^{\dagger}c_{-p\downarrow}^{\dagger}\right)|0\rangle,
	\end{eqnarray}
	where $u_p=\sqrt{\frac{1}{2}\left(1+\xi_{p}/\sqrt{\xi_{p}^2+\Delta^2}\right)}$ and $v_p=\sqrt{\frac{1}{2}\left(1-\xi_{p}/\sqrt{\xi_{p}^2+\Delta^2}\right)}$ are coefficients of the Bogoliubov transformation, $\xi_{p}$ and $\Delta$ are momentum and gap respectively.
	The fermionic twisting operator shifts the momenta of the two spin species in opposite directions, leading to
	$\bra{\text{G.S.}}\hat{U}^2\ket{\text{G.S.}} = \prod_{p}(u_{p+\frac{4\pi}{L}}u_{p}+v_{p+\frac{4\pi}{L}}v_{p})$.
	For any finite pairing gap $\Delta>0$, the coherence factors are smooth functions of momentum, giving~\cite{Append}
	\begin{eqnarray}\label{eq:bcs_overlap}
		\bra{\text{G.S.}}\hat{U}^2\ket{\text{G.S.}} = 1 - M\frac{8\pi}{L} + \mathscr{O}\left(\frac{1}{L^2}\right),
	\end{eqnarray}
	where $M = (2\pi/L) \sum_{p} \Delta^2\xi_{p}^{\prime2}/(\xi_{p}^2+\Delta^2)^2$. 
	This is consistent with\textit{Theorem 2} for a gapped fermionic state. 
	In contrast, when $\Delta \rightarrow 0$, the coherence factors become discontinuous at the Fermi points, 
	and the product is suppressed, signaling the gapless limit. 
	\begin{figure}[tbp]
		\centering
		\includegraphics[width=0.35\textwidth]{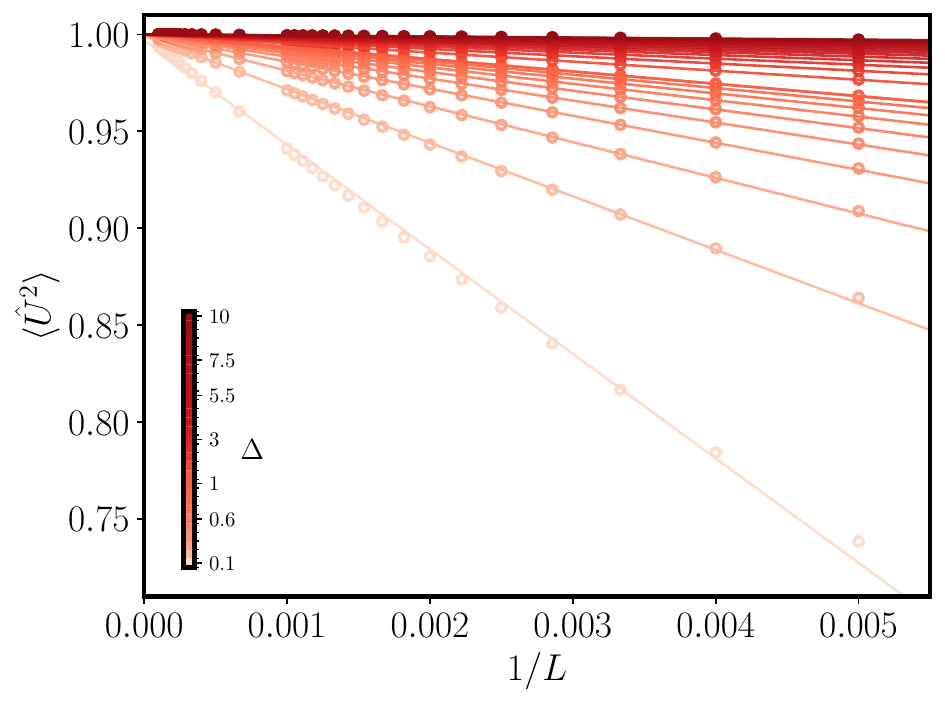}
		\caption{\justifying
			Results for $\langle\hat{U}^2\rangle$ in the BCS Hamiltonian. If we set $\Delta>0$, then $1-\braket{\hat{U}^2}\propto 1/L$ as shown by the different lines in the figure. For the given $L$, $\lim_{\Delta\to0}\braket{U}^2\to 0$ as shown in the vertical series of points in the figure.
		}
		\label{fig:MG}
	\end{figure}
	If $\Delta \to 0$, there exist $p$ such that $(u_{p+\frac{4\pi}{L}}u_{p}+v_{p+\frac{4\pi}{L}}v_{p})\to 0$,
	then the expectation will approach to zero for any given $L$, which signals the gapless limit.

\begin{figure}[tbp]
	\centering
	\subfigure{\includegraphics[width=0.2385\textwidth]{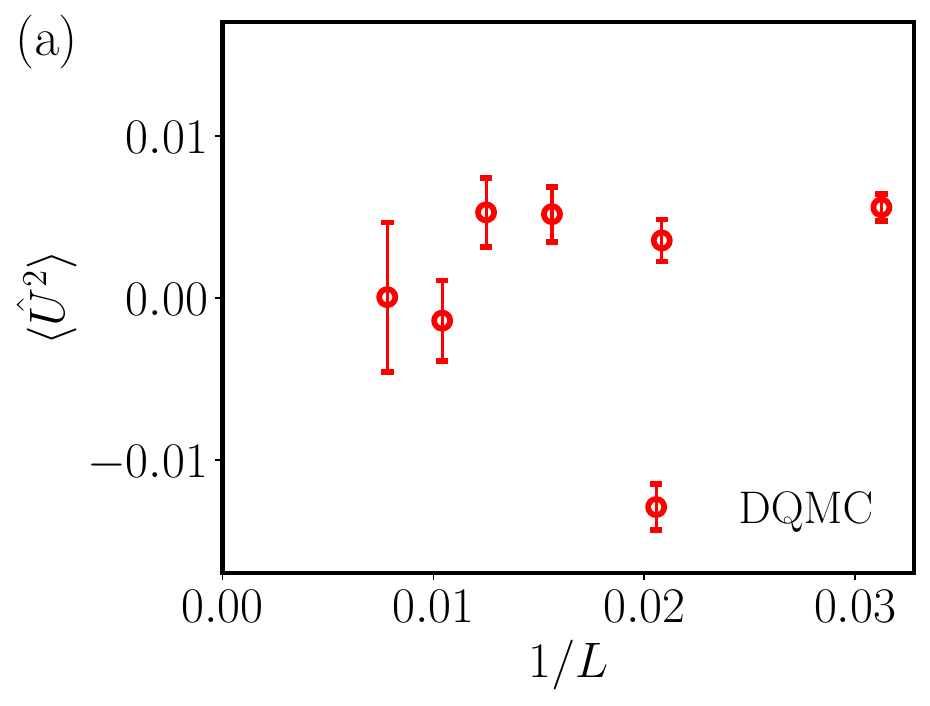}}
	\hfill
	\subfigure{\includegraphics[width=0.2385\textwidth]{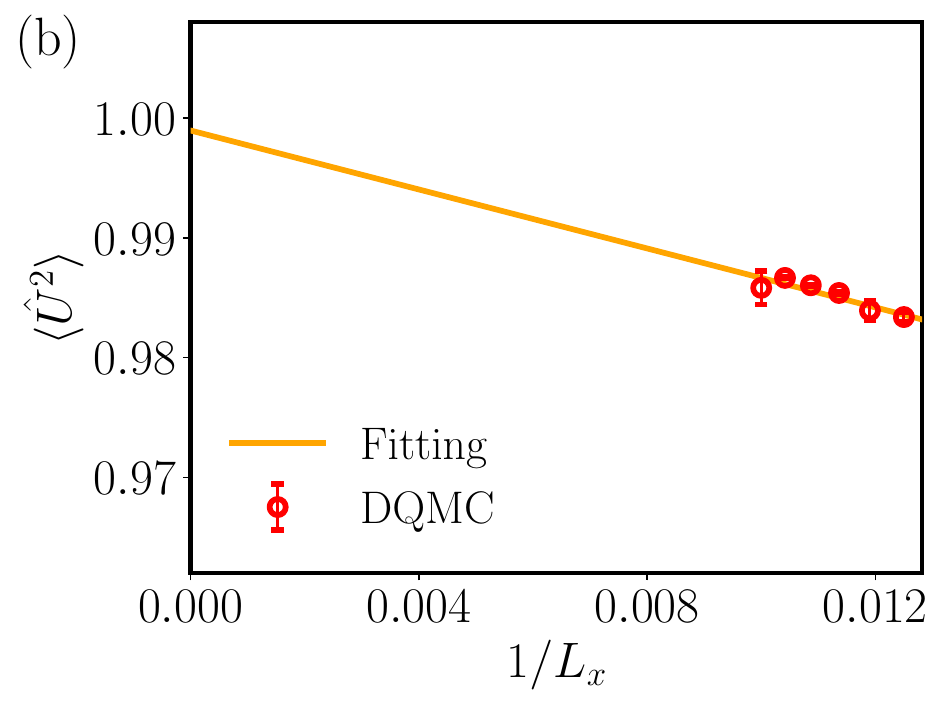}}
	\caption{\justifying
	Results for $\langle\hat{U}^2\rangle$ in interaction fermionic models. 
	(a) For the half-filled Hubbard chain, corresponding to the extended Hubbard model with $V=0$, $\langle \hat U^2\rangle$ remains close to zero for $L=32,\ldots,128$. 
	(b) For the half-filled two-leg Hubbard ladder with $U=4$, $t_{\parallel}=0.4$, and $t_{\perp}=1$, $\langle \hat U^2\rangle$ flows toward unity. 
	The solid line is a linear fit in $1/L_x$, with the intercept $0.999(3)$.
	}
	\label{fig:fermion_dqmc}
	\end{figure}

	We further test the fermion criterion in interacting models using DQMC. As a gapless example, we consider the one-dimensional extended Hubbard (EH) model~\cite{PhysRevLett.92.236401,PhysRevLett.99.216403}
	\begin{align}
	\mathcal{H}_{\text{EH}} &= -t\sum_{j,\sigma} \left(c^\dagger_{j\sigma}c_{j+1,\sigma}+\mathrm{h.c.}\right)  	+V\sum_j (n_j-1)(n_{j+1}-1)\nonumber \\
	&+U\sum_j\left(n_{j\uparrow}-\frac12\right)\left(n_{j\downarrow}-\frac12\right),
	\end{align}
	at half filling,
	where $n_{j\sigma}=c_{j\sigma}^\dagger c_{j\sigma}$ and $n_j=n_{j\uparrow}+n_{j\downarrow}$. 
	In the following calculation,
	we set $t=1$, $U=4$, and $V=0$, reducing the model to the repulsive Hubbard chain, whose spin sector remains gapless. 
	We choose the inverse temperature $\beta=L/4$ and compute the ground-state expectation value of the fermionic twisting operator. 
	As shown in Fig.~\ref{fig:fermion_dqmc}(a), $\langle \hat U^2\rangle$ remains close to zero up to $L=128$, consistent with the gapless nature of the model.

	As a gapped interacting example, we consider the half-filled two-leg Hubbard ladder~\cite{NOACK1996281},
	where $\lambda$ labels the ladder index:
	\begin{align}
	\mathcal{H}_{\mathrm{ladder}}&=-t_{\parallel}
	\sum_{\lambda=1,2}\sum_{x,\sigma}\left(c^\dagger_{x+1,\lambda,\sigma}c_{x,\lambda,\sigma}+\mathrm{h.c.}\right)\nonumber\\
	&-t_{\perp}\sum_{x,\sigma}\left(c^\dagger_{x,1,\sigma}c_{x,2,\sigma}+\mathrm{h.c.}\right)\nonumber\\
	&+U\sum_{\lambda=1,2}\sum_{x}\left(n_{x,\lambda,\uparrow}-\frac12\right)\left(n_{x,\lambda,\downarrow}-\frac12\right).
	\end{align}
	We use $U=4$, $t_{\perp}=1$, and $t_{\parallel}=0.4$, with PBCs along the ladder direction.  The system contains $2L_x$ sites, and the simulations are performed at $\beta=L_x/4$. 
	Figure~\ref{fig:fermion_dqmc}(b) shows that $\langle \hat U^2\rangle$ approaches unity as $L_x$ increases up to $L_x=100$.  A linear extrapolation in $1/L_x$ using the largest sizes gives the intercept $0.999(3)$, consistent with the expected thermodynamic value for a gapped fermionic state.

	\paragraph{SU($N$) generalizations.---}
	We extend the \textit{Theorems}~1 and 2 to the SU($N$) case, where each site can host a fermion with arbitrary SU($N$) ``spin''. 
	Its Hilbert space forms a linear representation of the SU(N).
We select an element of the Cartan subalgebra,
denoted by $\mathcal{S}^z$ in analog to spin-$z$ component in the case of SU(2),
represented by a matrix with elements $(\mathcal{S}^z)_{\alpha\beta}$.
We can construct spin operators using creation and annihilation operators, similar to the method used in the SU(2) case:
	\begin{eqnarray}
		\hat{\mathcal{S}}^z_{j} = \sum_{\alpha,\beta}c_{j,\alpha}^{\dagger}(\mathcal{S}^z)_{\alpha\beta}c_{j,\beta}.
	\end{eqnarray}
	The corresponding twisting operator is
	\begin{eqnarray}
		\hat{\mathcal{U}}= \exp \left( \frac{2\pi i}{L} \sum_{j = 1}^{L} j \hat{\mathcal{S}}^z_{j} \right),
	\end{eqnarray}
	where $2\pi$ is the smallest positive angle given in TABLE.~\ref{table 1}, so that the commutator of $\hat{\mathcal{U}}$ with the lattice translation is simply a phase times a global rotation as in Eq.~\eqref{TUT}.
	The remaining but crucial step is to find $\mathcal{R}$ such that:

\paragraph{Lemma 3.---}
	For any selected ${S}^z$ of the Cartan subalgebra of SU($N$), there exists $\mathcal{R}\in \text{SU}(N)$ such that
	\begin{eqnarray}
		\label{lemma1}	\sum_{k=0}^{n_R-1} (\mathcal{R}^\dagger)^k {\mathcal{S}}^z \mathcal{R}^k = 0,
	\end{eqnarray}
	where $n_R$ is the order of $\mathcal{R}$.
	
	\textit{Proof:}
	Since the Cartan subalgebra of SU($N$) is formed by traceless matrices, then the permutation matrix 
	\begin{eqnarray}\label{unity_alpha}
		\mathcal{R} = \exp(i\alpha)\underbrace{\begin{pmatrix}
			& 1\\
			\mathbb{I}_{N-1} &
		\end{pmatrix}}_{\equiv\mathcal{X}_N}
	\end{eqnarray}
	satisfies the above condition Eq.~\eqref{lemma1} with $n_R=N$, where $\exp(i\alpha)$ ensures the unity determinant and $\mathbb{I}_{N-1}$ is the $(N-1)\times(N-1)$ identity matrix. 
	For general ${\mathcal{S}}^z$, the choice of $\mathcal{R}$ is usually not unique. Table~\ref{table 1} lists several typical examples, where $\mathcal{Z}_N \equiv \begin{pmatrix}\mathbb{I}_{N-1}&\\&-1\end{pmatrix}$. 
		
	\begin{table*}
	\centering
	\begin{tabular*}{0.9\linewidth}{cccc|ccccc}
	\toprule
	Symmetry & ${\mathcal{S}}^z$ & $\mathcal{R}$ & &Symmetry & ${\mathcal{S}}^z$ & $\mathcal{R}$ && \\
	\midrule
	SU(2)& 
	$\mathcal{Z}_2$ &  $\mathcal{X}_2$ &&&$\begin{pmatrix}\mathcal{Z}_2&\\&\mathcal{Z}_3\end{pmatrix}$&$\begin{pmatrix}\mathcal{X}_2&\\&\mathcal{X}_3\end{pmatrix}$&$\mathcal{X}_5$&\\&&&&&&&&\\
	SU(3)&$\mathcal{Z}_3$ & $\mathcal{X}_3$&&SU(6)&
	$\mathcal{Z}_6$ &$\mathcal{X}_6$&&\\&&&&&&&&\\
	SU(4)&$\mathcal{Z}_4$&$\mathcal{X}_4$&&&$\begin{pmatrix}\mathcal{Z}_2&&\\&\mathcal{Z}_2&\\&&\mathcal{Z}_2\end{pmatrix}$&$\quad\begin{pmatrix}\mathcal{X}_2&&\\&\mathcal{X}_2&\\&&\mathcal{X}_2\end{pmatrix}\quad$&$\begin{pmatrix}\mathcal{X}_4&\\&\mathcal{X}_2\end{pmatrix}$&$\quad\mathcal{X}_6$\\&&&&&&&&\\
	&$\begin{pmatrix}\mathcal{Z}_2&\\&\mathcal{Z}_2\end{pmatrix}$&$\begin{pmatrix}\mathcal{X}_2&\\&\mathcal{X}_2\end{pmatrix}$&$\quad\mathcal{X}_4\quad$&&$\begin{pmatrix}\mathcal{Z}_4&\\&\mathcal{Z}_2\end{pmatrix}$&$\begin{pmatrix}\mathcal{X}_4&\\&\mathcal{X}_2\end{pmatrix}$&$\mathcal{X}_6$&\\&&&&&&&&\\
	SU(5)&$\mathcal{Z}_5$&$\mathcal{X}_5$&&&$\begin{pmatrix}\mathcal{Z}_3&\\&\mathcal{Z}_3\end{pmatrix}$&$\begin{pmatrix}\mathcal{X}_3&\\&\mathcal{X}_3\end{pmatrix}$&$\mathcal{X}_6$&\\
	\bottomrule
	
	\end{tabular*}
	\caption{\textbf{Suitable $\mathcal{R}$ for different $\mathcal{S}^z$}: $\mathcal{Z}_m \equiv \begin{pmatrix}\mathbb{I}_{m-1}&\\&-1\end{pmatrix}$, $\mathcal{X}_m\equiv\begin{pmatrix}
			& 1\\
			\mathbb{I}_{m-1} &
		\end{pmatrix}$.
		The phase factors such as in Eq.~\eqref{unity_alpha} to restore the unity determinant of $\mathcal{R}$'s are suppressed for simplicity.}\label{table 1}
	\end{table*}

	\textit{Theorem 4: Singlet criterion.---}
	Given a SU(N)-symmetric gapped fermionic chain Hamiltonian $\mathcal{H}$ with a finite ground-state degeneracy,
	any of its ground state(s) form a trivial representation of SU($N$).
	
	\textit{Proof: }
	We choose a basis of Cartan subalgebra of SU($N$) which are denoted as ${\mathcal{S}}^{z(i)}$ and $i=1,\cdots,N-1$,
	with its entries all equal to 0 except ${\mathcal{S}}^{z(i)}_{i,i}=1/2$ and ${\mathcal{S}}^{z(i)}_{i+1,i+1}=-1/2$.
	We also define $\mathcal{S}^{x,y(i)}$ where the non-zero elements of ${\mathcal{S}}^{x(i)}$ are ${\mathcal{S}}^{x(i)}_{i,i+1} = {\mathcal{S}}^{x(i)}_{i+1,i} = 1/2$, and the non-zero elements of ${\mathcal{S}}^{y(i)}$ are $-{\mathcal{S}}^{y(i)}_{i,i+1} = {\mathcal{S}}^{y(i)}_{i+1,i} = i/2$.
	We notice that the corresponding operators $\hat{\mathcal{S}}^{x,y,z(i)}$ forms a SU(2) subalgebra.
	Due to \textit{Theorem~1}, any of the ground state  must form a trivial representation of the such a SU(2) subalgebra. 
	Since $i$ is arbitrary,
	the entire basis $\{\hat{\mathcal{S}}^{z(i)}\}$ has zero eigenvalue on any ground state.
	Thus any gapped ground state $|\text{gs}\rangle$ must form a trivial representation of the entire SU($N$)~\cite{Hall:371445},
	so
	\begin{eqnarray}
	\mathcal{R}|\text{gs}\rangle=|\text{gs}\rangle.
	\end{eqnarray}
	This fact can be used to prove~\cite{Append}:

	\paragraph{Lemma 5.---}
	For any gapped spin-chain Hamiltonian $\mathcal{H}$ with a finite ground-state degeneracy possessing SU($N$) symmetry and any basis $\{|\text{gs}_j\rangle\}$ of its ground state(s),
	then we have: 
	\begin{eqnarray}
		\langle\text{gs}_i|\hat{\mathcal{V}}^{n_R}|\text{gs}_j\rangle = \langle\text{gs}_i|\hat{\mathcal{U}}^{n_R}|\text{gs}_j\rangle + \mathscr{O}(1/L),
	\end{eqnarray}
	where $\hat{\mathcal{V}}=\mathcal{R}\hat{\mathcal{U}}$.
	
	Then we can prove our main result:

	\paragraph{Theorem 6: $\langle\hat{\mathcal{U}}^M\rangle$-criterion.---} 
	Given an extensible spin-chain Hamiltonian $\mathcal{H}$ possessing SU($N$) symmetry,
	if there exists a normalized ground state, denoted as $|\text{gs}\rangle$,
	satisfying that
	\begin{eqnarray}
		\lim_{L\rightarrow\infty}\langle\text{gs}|\hat{\mathcal{U}}^M|\text{gs}\rangle\neq1 + \mathscr{O}(1/L),
	\end{eqnarray}
	then the Hamiltonian must be gapless,
	where $M=\gcd(\{n_R:\text{the order of all }\mathcal{R}\text{ satisfying Eq.~\eqref{lemma1}}\})$.
	
	\textit{Proof:} 
	We prove it by contradiction that the Hamiltonian is gapped. Taking the exponential mapping of Eq.\eqref{lemma1}, we obtain:
	\begin{eqnarray}
		1&=&(\mathcal{R}^{\dagger})^{n_R-1}\hat{\mathcal{U}}\mathcal{R}^{n_R-1}(\mathcal{R}^{\dagger})^{n_R-2}\cdots\hat{\mathcal{U}} \nonumber\\
		&=& (\mathcal{R}^{\dagger})^{n_R-1}\hat{\mathcal{U}}\mathcal{R}\hat{\mathcal{U}}\cdots \mathcal{R}\hat{\mathcal{U}} \nonumber = (\mathcal{R}\hat{\mathcal{U}})^{n_R} =\hat{\mathcal{V}}^{n_R},
	\end{eqnarray}
	so that we have by \textit{Lemma 5}:
	\begin{eqnarray}
		1 = \langle\text{gs}|\hat{\mathcal{V}}^{n_R}|\text{gs}\rangle = \langle\text{gs}|\hat{\mathcal{U}}^{n_R}|\text{gs}\rangle + \mathscr{O}(1/L).
	\end{eqnarray}
	Thus
	\begin{eqnarray}\label{u_R}
	\langle\text{gs}|\hat{\mathcal{U}}^{\pm n_R}|\text{gs}\rangle=1+\mathscr{O}(1/L).
	\end{eqnarray}
	Since $M$ can be written as
	\begin{eqnarray}
	M=\sum_R a_Rn_R,\,\,a_R\in\mathbb{Z},
	\end{eqnarray}
	then Eq.~\eqref{u_R} gives the most ideal estimate
	\begin{eqnarray}
			\langle\text{gs}|\hat{\mathcal{U}}^{M}|\text{gs}\rangle = 1 + \mathscr{O}(1/L).
	\end{eqnarray}

	\paragraph{Conclusions and discussions.---}
	%
	%
	In this work, we extend the twisting-operator criterion to spinful fermion chains and SU($N$)-symmetric spin chains.  
	For fermion chains with SU(2) spin-rotation symmetry, we prove that any gapped Hamiltonian with finite ground-state degeneracy satisfies $\langle \hat U^2\rangle=1+\mathscr{O}(1/L)$ in the thermodynamic limit.  
	We demonstrate this result using the BCS Hamiltonian and DQMC simulations of interacting Hubbard models, which distinguish the gapless Hubbard chain from the gapped two-leg Hubbard ladder. 
	For SU($N$)-symmetric spin chains, we prove that the gapped ground-state sector must form a trivial representation of SU($N$), and that $\langle \hat {\mathcal{U}}^M\rangle=1+\mathscr{O}(1/L)$ for a suitable positive integer $M$. 
	The extension of the present criterion to systems with reduced or discrete internal symmetries is left for future work.
	
	\paragraph{Acknowledgements.---}
	The authors thank Akira Furusaki and Linhao Li for helpful discussions, and Fo-Hong Wang and Tu Hong for useful discussions on DQMC simulations. 
	The DQMC simulations were performed using the \texttt{SmoQyDQMC.jl} package~\cite{SmoQyDQMC.jl}. 
	The computations in this Letter were run on the Siyuan-1 and $\pi$ 2.0 clusters supported by the Center for High Performance Computing at Shanghai Jiao Tong University.
	The work of Y.~Y. was supported by the National Key Research and Development Program of China (Grant No.~2024YFA1408303),
	the National Natural Science Foundation of China (Grant No.~12474157),
	the sponsorship from Yangyang Development Fund,
	and Xiaomi Young Scholars Program.

	\bibliographystyle{apsrev4-1}

%

	\newpage
	\onecolumngrid 
	\appendix
	\section{\large{} Supplemental Materials}
	
	\subsection{Analytical results of \texorpdfstring{$\langle G.S.|\hat{U}^2|G.S.\rangle$}{} for BCS Hamiltonian}
	In this section, we give a example about BCS Hamiltonian, Under the Mine-filed approximation, if BCS gap $\Delta >0$, there is a energy different between ground state and the first excited state, if $\Delta \to 0$, the system will trans to gapless phase. We will use twisting operator to show this transition.
	
	By using Bogoliubov transformation, we get BCS ground state:
	\begin{eqnarray}
		\ket{\text{G.S.}} = \prod_{p}\left(u_{p}+v_{p}c_{p\uparrow}^{\dagger}c_{-p\downarrow}^{\dagger}\right)\ket{0},
	\end{eqnarray}
	
	where
	
	\begin{eqnarray}
		u_{p} &=& \sqrt{\frac{1}{2}\left(1+\frac{\xi_{p}}{\sqrt{\xi_{p}^2+\Delta^2}}\right)}, \\
		v_{p} &=& \sqrt{\frac{1}{2}\left(1-\frac{\xi_{p}}{\sqrt{\xi_{p}^2+\Delta^2}}\right)}.
	\end{eqnarray}
	
	First, we can divide $\hat{U}$ into two parts: spin-up and spin-down
	\begin{eqnarray}
		\bra{\text{G.S.}}\hat{U}^2\ket{\text{G.S.}} &=& \bra{\text{G.S.}}\prod_{p}\hat{U}^2\left(u_{p}+v_{p}c_{p\uparrow}^{\dagger}c_{-p\downarrow}^{\dagger}\right)(\hat{U}^{\dagger})^2\ket{empty} \\ \nonumber
		&=&\bra{\text{G.S.}}\prod_{p}\left(u_{p}+v_{p}(\hat{U})^2c_{p\uparrow}^{\dagger}c_{-p\downarrow}^{\dagger}(\hat{U}^{\dagger})^2\right)\ket{empty} \\ \nonumber
		&=&\bra{\text{G.S.}}\prod_{p}\left(u_{p}+v_{p}\hat{U}^2c_{p\uparrow}^{\dagger}(\hat{U}^{\dagger})^2\hat{U}^2c_{-p\downarrow}^{\dagger}(\hat{U}^{\dagger})^2\right)\ket{empty} \\ \nonumber
		&=&\bra{\text{G.S.}}\prod_{p}\left(u_{p}+v_{p}\hat{U}_{\uparrow}^2c_{p\uparrow}^{\dagger}(\hat{U}_{\uparrow}^\dagger)^2\hat{U}_{\downarrow}^2c_{-p\downarrow}^{\dagger}(\hat{U}_{\downarrow}^{\dagger})^2\right)\ket{empty}. \\ \nonumber
	\end{eqnarray}
	
	Apply $\hat{U}$ to $c_p^{\dagger}$ will shift its momentum:
	
	\begin{eqnarray}
		\hat{U}_{\uparrow}c_{p\uparrow}^{\dagger}\hat{U}_{\uparrow}^\dagger &=& \exp\left(\frac{2\pi i}{L}\sum_{j=1}^{L}jc_{j\uparrow}^{\dagger}c_{j\uparrow}\right)c_{p\uparrow}^{\dagger} \exp\left(-\frac{2\pi i}{L}\sum_{j=1}^{L}jc_{j\uparrow}^{\dagger}c_{j\uparrow}\right) \\ \nonumber
		&=& \exp\left(\frac{2\pi i}{L}\sum_{j=1}^{L}jc_{j\uparrow}^{\dagger}c_{j\uparrow}\right)\frac{1}{\sqrt{L}}\sum_{m}e^{ipm}c_{m\uparrow}^{\dagger}\exp\left(-\frac{2\pi i}{L}\sum_{j=1}^{L}jc_{j\uparrow}^{\dagger}c_{j\uparrow}\right) \\ \nonumber
		&=&\frac{1}{\sqrt{L}}\sum_{m}e^{ipm}\exp\left(\frac{2\pi i}{L}mc_{m\uparrow}^{\dagger}c_{m\uparrow}\right)c_{m\uparrow}^{\dagger}\exp\left(-\frac{2\pi i}{L}mc_{m\uparrow}^{\dagger}c_{m\uparrow}\right) \\ \nonumber
		&=&\frac{1}{\sqrt{L}}\sum_{m}e^{ipm}\sum_{l=0}^{\infty}\frac{1}{l!}[\left(\frac{2\pi i}{L}mc_{m\uparrow}^{\dagger}c_{m\uparrow}\right)^{(l)},c_{m\uparrow}^{\dagger}]\\ \nonumber
		&=&\frac{1}{\sqrt{L}}\sum_{m}e^{ipm}\sum_{l=0}^{\infty}\frac{1}{l!}(\frac{2\pi i}{L}m)^{l}c_{m\uparrow}^{\dagger}\\ \nonumber
		&=&\frac{1}{\sqrt{L}}\sum_{m}e^{i(p+\frac{2\pi}{L})m}c_{m\uparrow}^{\dagger}\\ \nonumber
		&=&c_{p+\frac{2\pi}{L}\uparrow}^{\dagger}.\\ \nonumber
	\end{eqnarray}
	
	Similarly,
	
	\begin{eqnarray}
		\hat{U}_{\downarrow}c_{-p\downarrow}^{\dagger}\hat{U}_{\downarrow}^\dagger&=&c_{-(p+\frac{2\pi}{L})\downarrow}^{\dagger}.\\ \nonumber
	\end{eqnarray}
	
	So that
	\begin{eqnarray}
		\bra{\text{G.S.}}\hat{U}^2\ket{\text{G.S.}} &=& \bra{\text{G.S.}}\prod_{p}\left(u_{p}+v_{p}c_{p+\frac{4\pi}{L}\uparrow}^{\dagger}c_{-(p+\frac{4\pi}{L})\downarrow}^{\dagger}\right)\ket{0} \\ \nonumber
		&=&\prod_{p}(u_{p+\frac{4\pi}{L}}u_{p}+v_{p+\frac{4\pi}{L}}v_{p}),
	\end{eqnarray}
	
	and
	\begin{eqnarray}
		\bra{\text{G.S.}}\hat U^2\ket{\text{G.S.}}&=&\prod_{p}(u_{p+\frac{4\pi}{L}}u_{p}+v_{p+\frac{4\pi}{L}}v_{p}).
	\end{eqnarray}
	This result is consistent with the spin model, that is twisting operator can shift the crystal momentum of ground state.
	
	Now we discuss the change of the expectation of twisting operator as $\Delta$ approaches to zero.
	
	If $\Delta \ll 1$, by using Taylor expansion we have:
	\begin{eqnarray}
		u_{p+\frac{2\pi}{L}} &=& u_p + \frac{1}{2u_{p}}\left(\frac{\xi_{p}^{\prime}}{K} - \frac{\xi_{p}^2\xi_{p}^{\prime}}{K^{3}}\right)\frac{2\pi}{L} \\ \nonumber
		&+&  \frac{1}{2u_{p}}\left(-u_{p}^{\prime2} + \frac{\xi_{p}^{\prime\prime}}{2K} - \frac{3\xi_{p}\xi_{p}^{\prime2}+\xi_{p}^2\xi_{p}^{\prime\prime}}{2K^{3}}+\frac{3\xi_{p}^3\xi_{p}^{\prime2}}{2K^{5}}\right)\left(\frac{2\pi}{L}\right)^{2} + \dots, \\ \nonumber
		v_{p+\frac{2\pi}{L}} &=& v_p + \frac{1}{2v_{p}}\left(-\frac{\xi_{p}^{\prime}}{K} + \frac{\xi_{p}^2\xi_{p}^{\prime}}{K^{3}}\right)\frac{2\pi}{L} \\ \nonumber
		&+&  \frac{1}{2v_{p}}\left(-v_{p}^{\prime2} - \frac{\xi_{p}^{\prime\prime}}{2K} + \frac{3\xi_{p}\xi_{p}^{\prime2}+\xi_{p}^2\xi_{p}^{\prime\prime}}{2K^{3}}-\frac{3\xi_{p}^3\xi_{p}^{\prime2}}{2K^{5}}\right)\left(\frac{2\pi}{L}\right)^{2} + \dots, \\ \nonumber
	\end{eqnarray}
	where $K = \sqrt{\xi^{2}_{p}+\Delta^{2}}$, so that
	\begin{eqnarray}
		(u_{p+\frac{2\pi}{L}}u_{p}+v_{p+\frac{2\pi}{L}}v_{p}) &\approx& 1 - \frac{u_{p}^{\prime2}+v_{p}^{\prime2}}{2}\left(\frac{2\pi}{L}\right)^2 \\ \nonumber
		&=& 1 - \left(\frac{\xi_{p}^{\prime}}{K} - \frac{\xi_{p}^2\xi_{p}^{\prime}}{K^{3}}\right)^{2}(1+\frac{\xi^2_{p}}{\Delta^2})\left(\frac{2\pi}{L}\right)^2 \\ \nonumber
		&=& 1 - \frac{\Delta^2\xi_{p}^{\prime2}}{(\xi_{p}^2+\Delta^2)^2}\left(\frac{2\pi}{L}\right)^2.
	\end{eqnarray}
	
	If $\Delta > 0 $,
	there exist constant $M$ s.t.
	\begin{eqnarray}
		\frac{\Delta^2\xi_{p}^{\prime2}}{(\xi_{p}^2+\Delta^2)^2} < M.
	\end{eqnarray}
	
	We have
	\begin{eqnarray}
		\bra{\text{G.S.}}\hat{U}^2\ket{\text{G.S.}} &=& \prod_{p}(u_{p+\frac{4\pi}{L}}u_{p}+v_{p+\frac{4\pi}{L}}v_{p}) \\ \nonumber
		&=& \prod_{p}(1-M(\frac{4\pi}{L})^2) + \mathscr{O}(1/L^2) \\ \nonumber
		&=& 1 - M\frac{16\pi^2}{L} + \mathscr{O}(1/L^2).
	\end{eqnarray}
	Precisely,
	
	\begin{eqnarray}
		M = \sum_{p}\frac{\Delta^2\xi_{p}^{\prime2}}{(\xi_{p}^2+\Delta^2)^2}.
	\end{eqnarray}
	M can be calculated by integration when $L\to \infty$, It means if $
	\Delta > 0$, the expectation still approaches zero at a rate of $1/L$, and the coefficient depends on $\Delta$. 
	
	If $\Delta \to 0$:
	\begin{align}\left\{\begin{aligned}
			u_{p}=\left\{\begin{aligned}
				1,\quad \text{if} \quad \xi_{p}>0\\
				0,\quad \text{if} \quad \xi_{p}<0
			\end{aligned}\right.\\
			v_{p}=\left\{\begin{aligned}
				0,\quad \text{if} \quad \xi_{p}>0\\
				1,\quad \text{if} \quad \xi_{p}<0
			\end{aligned}\right.
		\end{aligned}\right.,\end{align}
	
	so that if $\xi_{p}$ and $\xi_{p+\frac{4\pi}{L}}>0$ have the same sign,
	$(u_{p}u_{p+\frac{4\pi}{L}}+v_{p}v_{p+\frac{4\pi}{L}}) \to 1$.
	if $\xi_{p}$ and $\xi_{p+\frac{4\pi}{L}}$ have different sign,
	for $\xi_{p}<0$:
	\begin{eqnarray}
		u_{p} &=& \frac{1}{2|\xi_{p}|}\Delta + \mathscr{O}(\Delta^2) \\ \nonumber
		u_{p+\frac{4\pi}{L}} &=& 1 - \frac{1}{2|\xi_{p+\frac{4\pi}{L}}|}\Delta + \mathscr{O}(\Delta^2) \\ \nonumber
		v_{p} &=& 1 - \frac{1}{2|\xi_{p}|}\Delta + \mathscr{O}(\Delta^2) \\ \nonumber
		v_{p+\frac{4\pi}{L}} &=& \frac{1}{2|\xi_{p+\frac{4\pi}{L}}|}\Delta + \mathscr{O}(\Delta^2).
	\end{eqnarray}
	
	Substituting above equation into $(u_{p}u_{p+\frac{4\pi}{L}}+v_{p}v_{p+\frac{4\pi}{L}})$ gives:
	\begin{eqnarray}
		(u_{p}u_{p+\frac{4\pi}{L}}+v_{p}v_{p+\frac{4\pi}{L}}) = \frac{1}{2}\left(\frac{1}{|\xi_{p}|}+\frac{1}{|\xi_{p+\frac{4\pi}{L}}|}\right)\Delta + \mathscr{O}(\Delta^2),
	\end{eqnarray}
	for  $\xi_{p}>0$,
	the result is similar. If the sign changes between $p$ and $p+\frac{2\pi}{L}$, there are 4 terms approach to zero
	\begin{eqnarray}
		\bra{\text{G.S.}}\hat{U}^2\ket{\text{G.S.}} = \frac{1}{16}\left(\frac{1}{|\xi_{p}|}+\frac{1}{|\xi_{p+\frac{4\pi}{L}}|}\right)^2\left(\frac{1}{|\xi_{p-\frac{2\pi}{L}}|}+\frac{1}{|\xi_{p+\frac{2\pi}{L}}|}\right)^2\Delta^4 + \mathscr{O}(\Delta^5),
	\end{eqnarray}
	this shows the expectation will approach to zero fro any given $L$, when $\Delta \to 0$, this means that the system trans to gapless phase.
	
	\subsection{Proof of Lemma~5}

	\textit{Proof: }$(n\equiv n_R)$
	Assume $\hat{\mathcal{U}}^{n}|\text{gs}\rangle = |\text{gs}^{(n)}\rangle + |\text{e}^{(n)}\rangle$ and $\hat{\mathcal{U}}^{\dagger}|\text{gs}\rangle = |\text{gs}^{\prime}\rangle + |\text{e}^{\prime}\rangle $ where $|\text{gs}^{(n)}\rangle$ represents the components on ground state sector and $|\text{e}^{(n)}\rangle$ represents the components on excited state sector, it has been shown that $|\text{e}^{(n)}\rangle$ has norm 
	$\sim \mathscr{O}(1/L)$~\cite{29j7-ltfn}.
	If $n=2$, because any ground state forms a trivial representation of SU($N$) according to \textit{Theorem 4},
	we have:
	\begin{eqnarray}
		\langle\text{gs}_i|\hat{\mathcal{V}}^2|\text{gs}_j\rangle = \langle\text{gs}_i|\hat{\mathcal{U}}\mathcal{R}\hat{\mathcal{U}}|\text{gs}_j\rangle.
	\end{eqnarray}
	Substitute it into the formula above to obtain:
	\begin{eqnarray}
		\langle\text{gs}_i|\hat{\mathcal{U}}\mathcal{R}\hat{\mathcal{U}}|\text{gs}_j\rangle &=& \langle\text{gs}^{\prime}_i|\text{gs}^{(1)}_j\rangle + \langle\text{e}^{\prime}_i|\mathcal{R}|\text{e}^{(1)}_j\rangle \nonumber \\
		&=& \langle\text{gs}^{\prime}_i|\text{gs}^{(1)}_j\rangle + \mathscr{O}(1/L) \nonumber \\
		&=& \langle\text{gs}_i|\hat{\mathcal{U}}^2|\text{gs}_j\rangle + \mathscr{O}(1/L).
	\end{eqnarray}
	Similarly, for $n = 3$
	\begin{eqnarray}
		\langle\text{gs}_i|\hat{\mathcal{V}}^3|\text{gs}_j\rangle &=& \langle\text{gs}^{\prime}_i|\hat{\mathcal{U}}|\text{gs}^{(1)}_j\rangle + \langle\text{e}^{\prime}_i|\mathcal{R}\hat{\mathcal{U}}\mathcal{R}|\text{e}^{(1)}_j\rangle \nonumber \\
		&=& \langle\text{gs}^{\prime}_i|\text{gs}^{(2)}_j\rangle + \mathscr{O}(1/L) \nonumber \\
		&=& \langle\text{gs}_i|\hat{\mathcal{U}}^3|\text{gs}_j\rangle + \mathscr{O}(1/L).
	\end{eqnarray}
	Repeating similar processes can prove that the lemma still holds for any given n.

\end{document}